\documentclass[twocolumn]{emulateapj}

\usepackage{amsmath}
\usepackage[title]{appendix}

\shorttitle{White Dwarfs in NGC 2323}
\shortauthors{Cummings et~al.}

\begin{document}

\title{Two Massive White Dwarfs from NGC 2323\\ 
and the Initial-Final Mass Relation for Progenitors of 4 to 6.5 M$_\odot$\textsuperscript{1}}


\author{Jeffrey D. Cummings\altaffilmark{2}, Jason S. Kalirai\altaffilmark{3,2},  P.-E. Tremblay\altaffilmark{4}, AND Enrico Ramirez-Ruiz\altaffilmark{5}}
\affil{}

\footnotetext[1]{Based on
observations with the W.M. Keck Observatory, which is operated as a scientific partnership
among the California Institute of Technology, the University of California, and NASA, was made 
possible by the generous financial support of the W.M. Keck Foundation.}

\altaffiltext{2}{Center for Astrophysical Sciences, Johns Hopkins University,
3400 N. Charles Street, Baltimore, MD 21218, USA; jcummi19@jhu.edu}
\altaffiltext{3}{Space Telescope Science Institute, 3700 San Martin Drive, Baltimore, MD 21218, USA;
jkalirai@stsci.edu}
\altaffiltext{4}{Department of Physics, University of Warwick, Coventry CV4 7AL, UK; 
P-E.Tremblay@warwick.ac.uk} 
\altaffiltext{5}{Department of Astronomy and Astrophysics, University of California,
Santa Cruz, CA 95064; enrico@ucolick.org} 

\begin{abstract}
We have observed a sample of 10 white dwarf candidates in the rich open cluster NGC 2323 (M50) with 
the Keck Low-Resolution Imaging Spectrometer.  The spectroscopy shows eight to be DA white dwarfs, 
with six of these having high S/N appropriate for our analysis.  Two of these white dwarfs are 
consistent with singly evolved cluster membership, and both are high mass $\sim$1.07 M$_\odot$, and 
give equivalent progenitor masses of 4.69 M$_\odot$.  To supplement these new high-mass white dwarfs 
and analyze the initial-final mass relation (IFMR), we have also looked at 30 white dwarfs
from publicly available data that are mostly all high-mass ($\gtrsim$0.9 M$_\odot$).  These original 
published data exhibited significant scatter, and to test if
this scatter is true or simply the result of systematics, we have uniformly analyzed the white 
dwarf spectra and have adopted thorough photometric techniques to derive uniform cluster parameters 
for their parent clusters.  The resulting IFMR scatter is significantly reduced, arguing that 
mass-loss rates are not stochastic in nature and that within the ranges of metallicity and
mass analyzed in this work mass loss is not highly sensitive to variations in 
metallicity.  Lastly, when adopting cluster ages based on Y$^2$ isochrones, the slope of the 
high-mass IFMR remains steep and consistent with that found from intermediate-mass white dwarfs, giving 
a linear IFMR from progenitor masses between 3 to 6.5 M$_\odot$.  In contrast, when adopting the
slightly younger cluster ages based on PARSEC isochrones, the high-mass IFMR has a moderate turnover 
near an initial mass of 4 M$_\odot$.

\end{abstract}

\section{Introduction}

Performing stellar archeology on white dwarfs, by far the most common stellar remnant,
provides valuable information for not only understanding stellar evolution and mass loss but
also galactic evolution.  One of the fundamental relations in the analysis of white dwarfs is 
the initial-final mass relation (hereafter IFMR), where the masses of white dwarfs are 
compared directly to the zero-age main sequence mass of their progenitors.  This semi-empirical 
relation is critical to our understanding of integrated mass-loss over the lifetime of a 
star and how it changes with stellar mass.  The IFMR has a variety of additional applications 
including predicting Type Ia supernovae rates (Pritchet et~al.\ 2008; Greggio 2010) and 
overall stellar feedback in galaxy models (Agertz \& Kravtsov 2014), interpreting the white 
dwarf luminosity function (Catal{\'a}n et~al.\ 2008), and providing a technique for measuring 
the age of the Galactic halo (Kalirai 2013).

Analysis of the IFMR began with Weidemann (1977), where it was shown that the models of the 
time greatly underestimated the observed stellar mass loss.  Subsequent work on the IFMR by a 
number of groups (see Weidemann 2000 for review) resulted in a broad but sparsely populated 
relation that showed a clear trend with higher-mass main sequence stars producing increasingly 
more massive white dwarfs.  In the past 15 years the amount of IFMR data has greatly increased 
(e.g., Claver et~al.\ 2001; Dobbie et~al.\ 2004, 2006a; Williams et~al.\ 2004; Kalirai et~al.\ 
2005; Liebert et~al.\ 2005; Williams \& Bolte 2007; Kalirai et~al.\ 2007; Kalirai et~al.\ 2008; 
Rubin et~al.\ 2008; Kalirai et~al.\ 2009; Williams et~al.\ 2009; Dobbie \& Baxter 2010; Dobbie 
et~al.\ 2012).  These newer data retain the general trend of the previous IFMR work, but the 
scatter in the data remains significant.  The source of this scatter may be attributable to 
several factors including the possible stochastic nature of mass loss, effects from
variation in metallicity or environment, or systematic differences between the studies.  One 
important systematic is the challenge in defining the ages of the clusters these white 
dwarfs belong to, which creates uncertainty in the derived lifetimes of their progenitor stars.
Cummings et~al.\ (2015; hereafter Paper I) began to analyze the important intermediate-mass IFMR (progenitor
masses of 3-4 M$_\odot$) from the rich NGC 2099 (M37).  This work strengthened the observational 
evidence that in this mass range the IFMR is steep, where the final white dwarf mass increases 
more rapidly with increasing progenitor mass.  Comparison to the rich population of comparable
mass white dwarfs in both the Hyades and Praesepe from Kalirai et~al.\ (2014) showed strongly 
consistent IFMRs.  This consistency also suggests that across this mass range the slightly 
metal-rich progenitor stars from the Hyades and Praesepe ([Fe/H]$\sim$0.15) have no 
significant increase in mass-loss rates compared to those in the solar metallicity NGC 2099.

Expanding beyond Paper I, we now look at the challenging higher-mass region (M$_{\rm initial}$ of 4--6.5
M$_\odot$) of the IFMR by focusing on white dwarfs in younger clusters.  While younger clusters 
do not provide a broad mass range of white dwarfs, they provide several important advantages.  
First, the highest-mass white dwarfs are the most compact and lowest luminosity, and because 
they form first they remain bright only in the youngest clusters ($<$ 200 Myr).  Second, the 
cooling rates are far more rapid in young and hot white dwarfs and, as a result, errors in
temperature lead to far smaller errors in both cooling age and luminosity.  Third, high-mass white 
dwarfs ($\gtrsim$0.9 M$_\odot$) may be prone to be ejected from their parent population clusters, either due to dynamical 
interactions or potential velocity kicks due to asymmetric mass loss during their formation (Fellhauer 
et~al.\ 2003; Tremblay et~al.\ 2012).  Therefore, the probability of finding high-mass white dwarfs 
still within their cluster population may decrease with age.  These three reasons are why younger 
clusters provide far more advantages in analyzing high-mass white dwarfs with the best precision.

In this paper we begin our analysis with the rich, young, and nearby cluster NGC 2323 (M50).  
Based on population analysis, Kalirai et~al.\ (2003) find that it is approximately three times 
as rich as the Pleiades, making it an excellent environment to search for the rare remnants 
of higher-mass stars.  Additionally, to expand our sample we self-consistently
reanalyze publicly available data on all published high-mass white dwarfs from clusters 
(Pleiades [M45], NGC 2516, NGC 2287, NGC 3532, and NGC 2168 [M35]) and Sirius B.  To further limit 
systematics, we also self-consistently analyze the cluster parameters for all of the parent 
cluster populations based on published high-quality UBV photometry.  To look at the broader 
picture, we connect these high-mass data to the moderate-mass data from Paper I and further 
analyze the broader characteristics of the IFMR.

The structure of this paper is as follows, in Section 2 we discuss the spectroscopic white 
dwarf observations of NGC 2323, the publicly available data we have used, and our reduction and
analysis techniques.  In Section 3 we discuss the UBV photometry based cluster parameters for 
our six open clusters being analyzed.  In Section 4 we discuss the cluster membership of our 
white dwarf candidates in NGC 2323.  In Section 5 we discuss the high-mass IFMR and compare to 
the intermediate-mass (3--4 M$_\odot$ progenitors) IFMR from Paper I.  In Section 6 we summarize 
our study.

\section{Observations, Reductions \& Analysis}

Based on the deep BV photometric observations of NGC 2323 (Kalirai et~al.\ 2003) with the 
Canada-France-Hawaii telescope and the CFH12K mosaic camera, a sample of white dwarf candidates 
in NGC 2323 were spectroscopically observed at Keck I using the Low Resolution Imaging Spectrometer 
(LRIS; Oke et~al.\ 1995).  In total, 10 of these candidates had sufficient signal to properly
analyze their characteristics.  The 400/3400 grism was used with 1" slits giving a spectral 
resolution of $\sim$6 \AA, which provides us the wavelength coverage of $\sim$3000 to 5750 
\AA\, and a series of 5 Balmer lines (H$\beta$, H$\gamma$, H$\delta$, H$\epsilon$, and H8).  
These observations were performed on 2008 December 23 and 24, on 2011 December 27 and 28, and 
on 2015 February 19.  For spectral flux calibration, flux standards were observed each night.  

Three independent LRIS masks were used in 2008 to observe white dwarfs candidates from NGC 
2323.  Mask 1 was observed for 40 minutes, Mask 2 was observed for 40 minutes, and Mask 3 was 
observed for 2 hours and 40 minutes.  In 2011, individual longslit observations were performed 
for five different white dwarf candidates in NGC 2323 ranging from 20 minutes to 1 hour and 50 
minutes.  Lastly, in 2015 an additional 70 minutes was acquired on WD11, which is a white dwarf 
candidate of interest that only had 20 minutes of observation in 2011.  We have reduced and flux 
calibrated our LRIS observations using the IDL based XIDL pipeline\footnote[6]{Available at 
http://www.ucolick.org/$\sim$xavier/IDL/}.  Of our total observed sample of 10 white dwarfs 
candidates, eight are DA white dwarfs and WD23 and WD38 have no clear spectral features.  

To provide additional high-mass white dwarfs for comparison we have taken from the VLT
Archive the observations of seven white dwarf members of NGC 3532, three white dwarf members
of NGC 2287, and four white dwarf members of NGC 2516 (Based on observations made with
ESO telescopes under Program IDs: 079.D-0490(A); 080.D-0654(A); 084.D-1097(A); PI: Dobbie).  
These observations were performed with FORS1 and FORS2 using the 600B grism (Appenzeller et~al.\ 
1998) giving comparable spectral resolution to our LRIS observations of $\sim$6 \AA.  Analysis 
of the parameters and membership for the white dwarfs 
of NGC 2287, NGC 2516, and NGC 3532 were originally published in Dobbie et~al.\ (2009; 2012).  
Additionally, from the Keck Archive we have taken the observations of 11 white dwarf members
of NGC 6128 (Program ID: U49L-2002B; U60L-2004A; U15L-2004B; U18L-2005B; PI: Bolte).
Similar to our NGC 2323 observations, these observations were performed using LRIS with a 
majority of them using the 400/3400 grism and 1" slits, giving the same 
characteristics to our data.  Analysis of the parameters and membership for the white dwarfs 
of NGC 6128 were original published in Williams et~al.\ (2009).  

We have performed our 
own reductions and analyses of these data from both the VLT and Keck, but we do not redetermine 
their membership status in this paper.  The VLT data were reduced using the standard IRAF 
techniques for reduction of longslit data, while the Keck data were reduced using the same XIDL 
pipeline used to analyze our NGC 2323 data.  As a test for our VLT data reduction, we were provided 
with the published spectrum of white dwarf J0646-203 (NGC 2287-4) (P.D. Dobbie; private 
communication 2014), and we found that there are no meaningful differences in our reduced spectra
of the same data.  Hence, there are no systematics caused by our spectral reduction techniques, 
and the systematic differences between our parameters and those presented in Dobbie et~al.\ 
(2012) are due to differences in our applied white dwarf models and fitting techniques.

\begin{center}
\begin{deluxetable*}{l c c c c c c c c c c}
\multicolumn{11}{c}%
{{\bfseries \tablename\ \thetable{} - NGC 2323 White Dwarf Initial and Final Parameters}} \\
\hline
ID&T$_{\rm eff}$&log g&M$_{WD}$   &M$_V$&t$_{cool}$& Y$^2$ M$_i$ & PARSEC M$_i$ &M$_{i120}$ &M$_{i160}$&S/N\\
  &(K)      &     &(M$_\odot$)& &(Myr)     &(M$_\odot$)&(M$_\odot$)&(M$_\odot$)&(M$_\odot$)&\\
\hline
\multicolumn{11}{l}{{Likely White Dwarf Cluster Members}}  \\
\hline
NGC 2323-WD10  & 52800$\pm$1350 & 8.68$\pm$0.09 & 1.068$\pm$0.045 & 10.36$\pm$0.19 & 1.6$^{+1.2}_{-0.6}$ & 4.69$^{+0.01}_{-0.01}$ & 5.07$^{+0.02}_{-0.02}$  &  4.98 & 4.45  & 85\\
NGC 2323-WD11  & 54100$\pm$1000 & 8.69$\pm$0.07 & 1.075$\pm$0.032 & 10.36$\pm$0.13 & 1.3$^{+0.6}_{-0.4}$ & 4.69$^{+0.01}_{-0.01}$ & 5.07$^{+0.02}_{-0.01}$  &  4.98 & 4.45  & 130\\
\hline
\multicolumn{11}{l}{{White Dwarfs Inconsistent with Single Star Membership}}  \\
\hline
NGC 2323-WD21  & 18200$\pm$850  & 8.26$\pm$0.15 & 0.779$\pm$0.096 & 11.33$\pm$0.25 & 170$^{+60}_{-48}$  & -- & --  &  --  & --  & 28\\
NGC 2323-WD7   & 16800$\pm$250  & 7.90$\pm$0.05 & 0.559$\pm$0.024 & 10.92$\pm$0.07 & 112$^{+12}_{-11}$  & -- & --  &  --  & --  &122\\
NGC 2323-WD17  & 19800$\pm$300  & 8.12$\pm$0.05 & 0.694$\pm$0.028 & 10.97$\pm$0.07 &  96$^{+12}_{-11}$  & -- & --  &  --  & --  &111\\
NGC 2323-WD12  & 17100$\pm$400  & 7.88$\pm$0.07 & 0.550$\pm$0.037 & 10.87$\pm$0.11 & 101$^{+17}_{-15}$  & -- & --  &  --  & --  & 60\\
\hline
\multicolumn{11}{l}{{Low Signal to Noise White Dwarfs}}  \\
\hline
NGC 2323-WD22  & 24400$\pm$1550 & 8.08$\pm$0.22 & 0.681$\pm$0.128 & 10.53$\pm$0.36 &  33$^{+36}_{-16}$  & -- & --  &  --  &  --  & 13\\
NGC 2323-WD30  & 13400$\pm$900  & 8.04$\pm$0.18 & 0.629$\pm$0.106 & 11.52$\pm$0.28 & 290$^{+108}_{-82}$ & -- & --  &  --  &  --  & 23\\
\hline
\end{deluxetable*}
\end{center}

\vspace{-1cm}
For the high-mass LB 1497, from the Pleiades, and Sirius B we have taken the T$_{\rm eff}$ 
and log g parameters from Gianninas et~al.\ (2011).  We also have taken these parameters from
Gianninas et~al.\ (2011) for the supermassive GD50 and PG 0136+251 white dwarfs, where Dobbie
et~al.\ (2006b) used three-dimensional space velocities to argue that GD50's progenitor was related
to and coeval with the Pleiades cluster.  Similarly, but based on only proper motions, Dobbie
et~al.\ (2006b) argued that PG 0136+251's progenitor is likely consistent with coeval formation 
with the Pleiades.  Gianninas et~al.\ (2011) use white dwarf atmospheric models and fitting 
techniques equivalent to ours, and we adopt their T$_{\rm eff}$ and log g in our 
analysis of these four white dwarfs.

For our spectroscopic analysis we adopted the same analysis techniques as those described
in Paper I.  In brief, we used the recent white dwarf spectroscopic models of Tremblay et~al.\ 
(2011) with the Stark profiles of Tremblay \& Bergeron (2009), and the automated fitting techniques 
described by Bergeron et~al.\ (1992) to fit our Balmer line spectra and derive T$_{\rm eff}$ 
and log g.  However, for our derived parameters (mass, luminosity, and cooling age) we expand 
upon the methods of Paper I because our current sample has a far broader mass range.  
For deriving the parameters for white dwarfs of mass less than 1.10 M$_\odot$ we applied our 
T$_{\rm eff}$ and log g to the cooling models for a carbon/oxygen (CO) core composition with a 
thick hydrogen layer by Fontaine et~al.\ (2001).  For the highest-mass white dwarfs ($>$ 1.1
M$_\odot$) we derived the parameters based on the oxygen/neon (ONe) core 
models of Althaus et~al.\ (2005; 2007).  

\begin{figure}[!ht]
\begin{center}
\includegraphics[clip, scale=0.43]{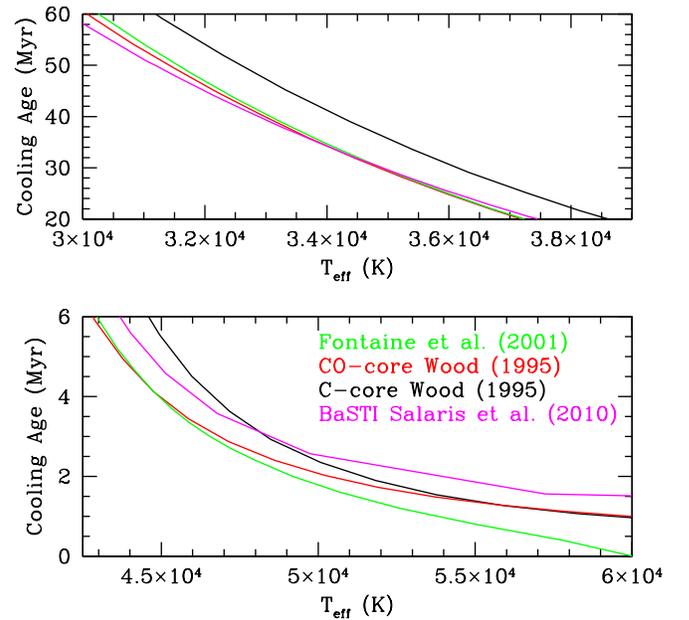}
\end{center}
\vspace{-0.4cm}
\caption{A comparison at 1 M$_\odot$ of the cooling rates for the several white dwarf models 
that we have discussed, in addition to the BaSTI CO cooling models from Salaris et~al.\ (2010).  
In the lower panel, the hottest temperatures are shown and the systematics are always $\sim$1.5 Myr 
or less in magnitude, which is not concerning.
At cooler temperatures, all models other than the carbon core models of Wood (1995) are consistent.
This is consistent with the systematic effects of differing core composition, which grow as white
dwarfs cool further.}
\end{figure}

A discussion of our adopted models is warranted.  First, we do not adopt the more recent white dwarf 
atmospheric models of Tremblay et~al.\ (2013) because those focus only on 3D modeling of convective 
atmospheres ($\lesssim$14,000 K), and our current analysis looks hot fully radiative white 
dwarfs.  For our cooling models, while the Fontaine et~al. (2001) models are widely used, we should 
acknowledge two limitations they have.  First, they assume a 50/50 carbon and oxygen core composition, 
which based on full stellar evolutionary models is not accurate (e.g., Romero et~al.\ 2013).  The 
effect of this on the calculated cooling ages can be important, but for the relatively young white 
dwarfs we are analyzing this effect remains small.  Second, these cooling models do not begin at the 
tip of the asymptotic giant branch (hereafter AGB) and instead begin at a T$_{\rm eff}$ of $\sim$60,000 
K.  In comparison, the widely used Wood (1995) CO cooling models do begin at the tip of the AGB 
and they do not adopt a simple 50/50 carbon and oxygen core composition\footnote[7]{Equation 1 of Wood (1995)
gives their C/O ratio relation.}.  Unfortunately, their CO core models 
have an upper mass limit of 1.0 M$_\odot$, which limits their application in our analysis but they provide
an important test for systematics.  In Figure 1 we compare the cooling ages from the Fontaine et~al.\ (2001) 
CO core models at 1.0 M$_\odot$ to both the CO and pure-carbon core models of Wood (1995) and the BaSTI CO 
models from Salaris et~al.\ (2010).  The BaSTI CO cooling models adopted C/O ratio profiles based 
on the BaSTI scaled solar stellar evolution models.  The lower panel of Figure 1 shows that at high T$_{\rm 
eff}$$>$45,000 K all of the differences between these models never result in cooling age differences of more 
than 1.5 Myr.  The upper panel of Figure 1 similarly shows that at cooler temperatures 30,000$<$T$_{\rm eff}$$<$40,000 
K the CO models, irrespective of their adopted C/O ratios or starting points, all give strong agreement 
in cooling ages, but the systematic effects introduced from pure carbon core models ($\sim$10 Myr) are now
clearly seen.

For ultramassive white dwarfs, the mass at which they transition to ONe white dwarfs remains 
uncertain, and it also likely depends on metallicity (Doherty et~al. 2015).  Here we have adopted a 
somewhat conservative 1.10 M$_\odot$, but we note that the models of Garcia-Berro et~al.\ (1997) argue that it 
may be as low as 1.05 M$_\odot$.  Reassuringly, in this mass range the CO and ONe cooling ages at constant mass are 
consistent for such young white dwarfs, but for these white dwarfs the dependence of the mass-radius 
relationship on core-composition is very important.  For example, applying the gravities of white dwarfs in 
the mass range of 1.05 to 1.10 M$_\odot$ (based on CO-core models) to the ONe models derives masses 
$\sim$0.05 M$_\odot$ lower and places them all below 1.05 M$_\odot$.  Therefore, it is appropriate
to adopt 1.10 M$_\odot$ as the transition mass.

\begin{figure*}[!ht]
\begin{center}
\includegraphics[clip, scale=0.9]{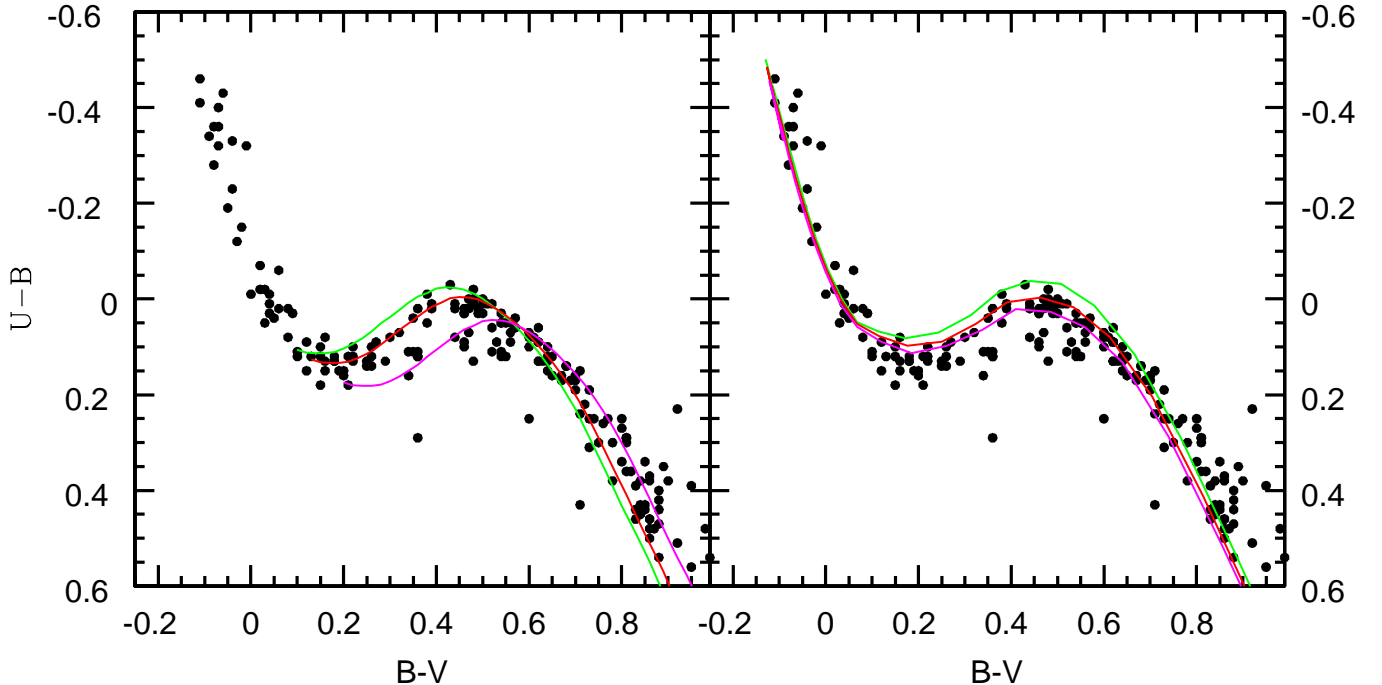}
\end{center}
\vspace{-0.4cm}
\caption{Color-color analysis of the Pleiades.  In the left panel we have plotted three Hyades
fiducial curves of [Fe/H]=0.01 (Z=0.0185) with differing reddening (Green E(B-V)=0.0, Red 
E(B-V)=0.03, Magenta E(B-V)=0.1) to the Pleiades data.  This finds that near B-V=0.6, where
all three reddening curves intersect, the photometry is not dependent on reddening.  Therefore,
the photometric metallicity is first matched by fitting this region to the data.  Of the three
curves a reddening of E(B-V)=0.03 (red curve) provides the best fit in the region spanning B-V 
of 0.1-0.5.  In the right panel we have focused on the hottest stars (B-V$<$0.0) and plot
three Y$^2$ isochrone curves of E(B-V)=0.03 with differing metallicity (Green Z=0.0125, Red 
Z=0.0185, and Magenta Z=0.0245).  These models extend farther into the blue, and this demonstrates 
that while the other regions of the diagram are sensitive to metallicity, these higher-mass star 
colors are only dependent on reddening.}
\end{figure*}

In Table 1 we present our white dwarf parameters for the eight DA white dwarfs from NGC 2323.
For clarity in Table 1 we have distinguished between members and nonmembers, where membership 
is based on our comparisons of model based and observed photometry in addition to 
comparisons of cooling ages to the cluster age (see our detailed discussion 
of membership in Section 4).  Additionally, we have considered S/N (per 
resolution element) and the resulting errors given in Table 1.  The WD22 and WD30 spectra have low 
S/N and mass errors greater than 0.1 M$_\odot$.  Therefore, their parameters are presented for 
reference but have been cut from our final analysis.  This is because their membership 
determinations are unreliable, but their low masses do suggest they are field white dwarfs.

\section{Cluster Parameters}

With IFMR analysis, the parameters of the star clusters are as critical as the white
dwarfs themselves.  This is particularly true for
the highest-mass white dwarfs, where the derived masses of their progenitors change
rapidly with evolutionary time.  Uniform photometric data sets are not available for these
six clusters.  But in Paper I we showed that for NGC 2099, the adopted isochrones and fitting
techniques had as large, if not larger, of an effect on its cluster parameters as
did any systematics between the cluster's different photometry sets.  
This is best illustrated by the systematic effects on derived cluster ages, where the 
Yi et~al.\ (2001; hereafter Y$^2$) and the Ventura et~al.\ (1998) isochrones both 
gave our final adopted age of 520 Myr for NGC 2099.  The PARSEC version 1.2S isochrones 
(Bressan et~al.\ 2012) derived a comparable age of 540 Myr, and lastly the Girardi et~al.\ 
(2000) and Bertelli et~al.\ (1994) isochrones both gave significantly younger ages of 445 Myr.
Between these isochrone sets there is nearly a 100 Myr range of derived ages when using 
identical photometry, and further systematics can be introduced based on how the isochrones 
are fit to the data.  

In this paper we have redetermined as uniformly as possible the reddenings, distance
moduli, and ages for these six clusters using available high-quality UBV photometry that 
covers up to the full turnoff.  The Y$^2$ isochrones provide our final adopted cluster ages, 
but to broaden our results we also determine ages with the 
PARSEC isochrones.  These two isochrones give only slightly different cluster ages, but 
in these younger clusters the masses of the progenitor stars have a far more significant 
dependence on evolutionary time, and even a 20 Myr systematic has a significant effect 
on our results.  

\vspace{0.3cm}
\subsection{Color-Color Analysis}

In our cluster photometric analysis we first make use
of color-color diagrams (B-V vs U-B), which provide direct photometric information on the cluster
reddening.  The photometric metallicity can also be derived but it is quite sensitive
to systematics in U magnitude, a concern considering the typically more complex 
standardization process for U magnitudes and the varying sources of our photometry.  Therefore, 
in the case of the Pleiades, NGC 2168, and NGC 2516 we consider more detailed spectroscopic 
metallicities, but for NGC 2287, NGC 2323, and NGC 3532 we will simply adopt solar 
metallicity.  However, we note that these adopted metallicities do show strong consistency
with the observed photometry.  Our color-color analysis adopts two techniques, the first is 
semi-empirical and based on the Hyades fiducial and the second is based directly on the Y$^2$
isochrones, which reach to higher masses than those available in the Hyades fiducial.  For 
both methods the reddening relation adopted is that of 
Cardelli et~al.\ (1989) and the metallicity correction is based on that of the Y$^2$ isochrones.  
The methods using the Hyades fiducial have been developed in Deliyannis et~al.\ (in prep), where
the fiducial was derived from single-star cluster members (see Perryman et~al.\ 1998).  The Hyades UBV 
photometry of Johnson \& Knuckles (1955) was adopted with a cluster [Fe/H] of +0.15 and E(B-V) of 0.

The Pleiades provides a good example for our color-color analysis techniques.  In the left panel of 
Figure 2 we have plotted the photoelectric UBV photometry from Johnson \& Mitchell (1958) with 
several reddening curves based on the Hyades fiducial, and we have applied a metallicity of 
[Fe/H]=0.01 (Z=0.0185) to match the reddening insensitive region where all of the reddening curves 
intersect near B-V of 0.6.  Fitting by eye the blueward color-color trend we find that a 
reddening curve of E(B-V)=0.03$\pm$0.02 matches the Pleiades UBV photometry the best.  Both this 
reddening and metallicity are consistent with the typically derived values and spectroscopic analyses 
of the Pleiades (e.g., [Fe/H]=0.01$\pm$0.02 Schuler et~al.\ 2010; [Fe/H]=0.03$\pm$0.02$\pm$0.05 
Soderblom et~al.\ 2009) We also note that the Hyades fiducial ends at B-V$\sim$0.1, where the older 
Hyades turnoff occurs.

In the right panel of Figure 2 we fit the higher-mass stars bluer than B-V=0.0 with three 
different 135 Myr Y$^2$ isochrones of differing metallicity.  All three metallicities fit a
reddening of E(B-V)=0.03 in these bluest stars.  This demonstrates that these higher-mass 
stars create a nearly linear 
trend that is insensitive to variations in metallicity.  We note that the position of this blue linear trend 
is also insensitive to cluster age, where as we look at older clusters the trend only shortens 
in length and does not shift its position.  Therefore, this linear trend's position
provides a reliable reddening measurement independent of all other cluster parameters.

\begin{figure*}[!ht]
\begin{center}
\includegraphics[clip, scale=0.91]{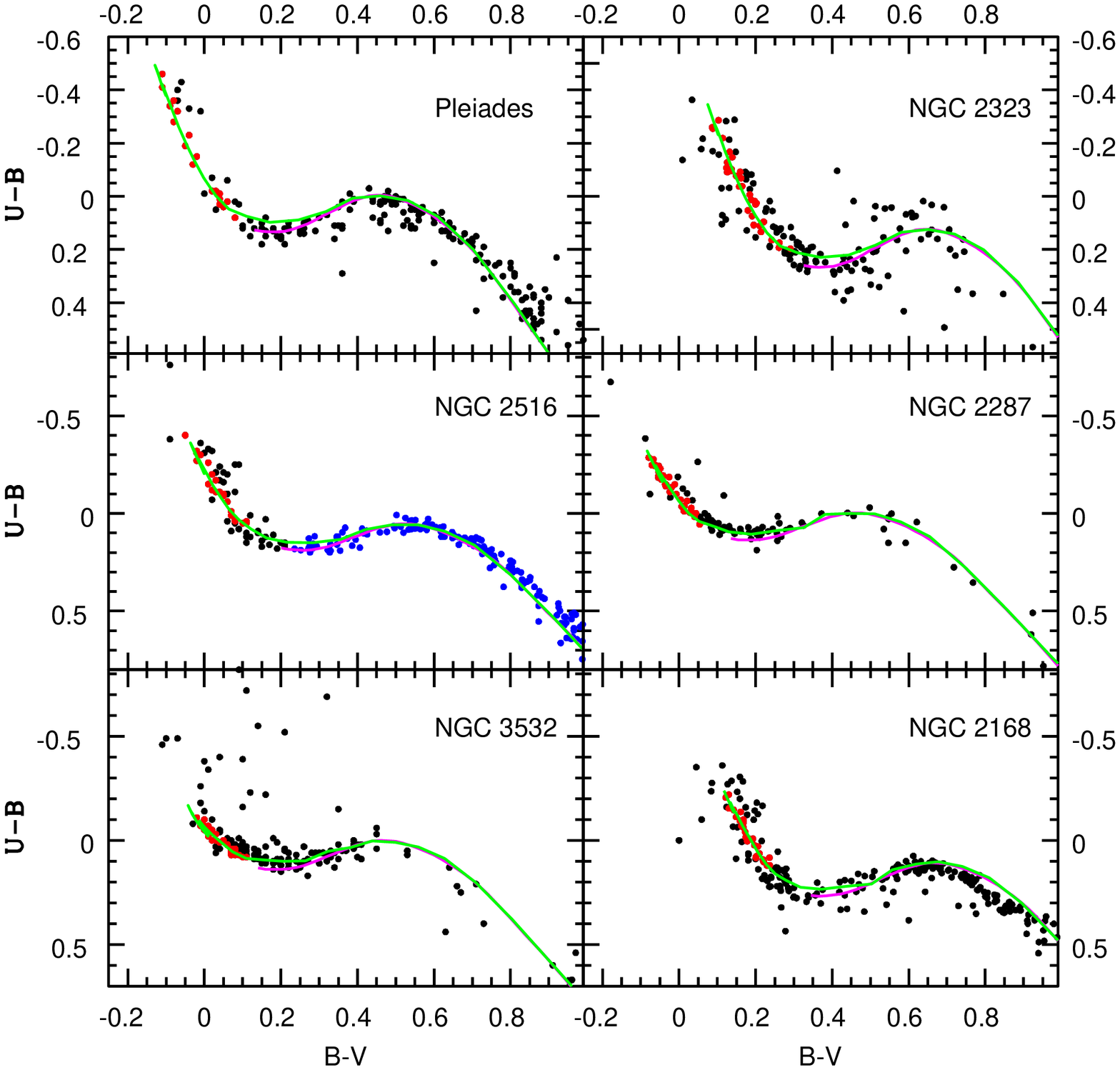}
\end{center}
\vspace{-0.4cm}
\caption{Our color-color analysis of all six of our clusters.  The full data sets are displayed
in black, with the NGC 2516 data being supplemented by additional data in blue.  Our Y$^2$
color-color fits are shown in green, and the comparable Hyades-fiducial based fits adopting the
same metallicities and reddenings are shown in magenta.  We use the Y$^2$ relations to identify
the most reliable turnoff stars shown in red, which we will use in our cluster age analysis.  See
Table 2 and the text for our photometric sources.}
\end{figure*}

In the upper left panel of Figure 3 we compare directly our Hyades fiducial
fit and our Y$^2$ fit for the Pleiades.  There are systematic differences that 
are noted in the B-V range of 0.1-0.3, where it appears that the Y$^2$ isochrones
are too blue in U-B relative to both the data and the Hyades fiducial.  However, when adopting 
a Y$^2$ isochrone with an age consistent to the Hyades (650 Myr; not shown) the isochrone is
nearly identical to that of the Hyades fiducial, so at these younger ages the isochrones
appear to overestimate the U flux at this intermediate color range.  At all other colors 
this does not seem to be a concern because, reassuringly, both the Hyades fiducial and the 
Y$^2$ isochrones find the same reddening in their respective regions, and they also agree 
in the region sensitive to metallicity (B-V$\sim$0.6) and redder.  In regard to the PARSEC isochrones, 
they are not independently considered in our color-color analysis because their U-B colors 
do not fit the observations.

A key advantage to color-color analysis is that it can be used to clean young cluster turnoffs.
Dating younger clusters is prone to several difficulties, including that turnoffs are relatively
sparse even in the richest clusters and that these higher-mass turnoff stars 
have high binarity fraction (Kouwenhoven et al. 2007).  Additionally, many higher-mass 
stars fall into the peculiar groups of blue stragglers or Be stars.  For example, Mermilliod 
1982a, 1982b, and Ahumada \& Lapasset (2007) found that several of the brightest 
stars in the clusters being analyzed in this paper are blue stragglers that are far too blue or too 
bright, if not both, for their age.  Lu et~al.\ (2011) modeled the formation of blue stragglers on the 
short time scales necessary in these young clusters and found that they can rapidly be 
created through binary mass transfer.  Identifying these peculiar stars can 
greatly improve the fit of the turnoff ages of these younger clusters.  Mermilliod 1982b do find 
that when plotted in color-color space Be stars deviate from the approximately linear trend of the 
``normal" high-mass stars.  Additionally, several of our clusters may have variable reddening, and
fitting this high-mass linear trend identifies the richest group of cluster stars with 
consistent reddening.   In the Pleiades shown in the upper-left panel of Figure 3, we mark in red
the high-mass stars consistent with the trend, and several stars deviate from this trend 
that are likely peculiar.  However, we do acknowledge that the brightest star in the Pleiades, Alcyone,
does not deviate from this trend.  Alcyone is a multiple system and has several peculiar 
characteristics like spectral emission and rapid rotation (Hoffleit \& Jaschek 1991) and is 
commonly referred to as both a blue straggler and Be dwarf.  This suggests that while this 
color-color method does identify many problematic stars, it does not remove all peculiar stars.

In the upper right panel of Figure 3 we have similarly plotted the NGC 2323 UBV photometry from Claria 
et~al.\ (1998).  We fit the bluest stars and see that independent of an assumed metallicity, 
we find a large reddening of E(B-V)=0.23$\pm$0.06 at (B-V)$_0$=0, where we deem this reddening
large enough to account for the color dependence of reddening (see Fernie 
1963 and our discussion in Paper I).  When we have adopted a color dependent reddening we 
will define the reddening at (B-V)$_0$=0.  Again, our selected final turnoff stars are shown 
in red.

\begin{figure*}[!ht]
\begin{center}
\includegraphics[clip, scale=0.92]{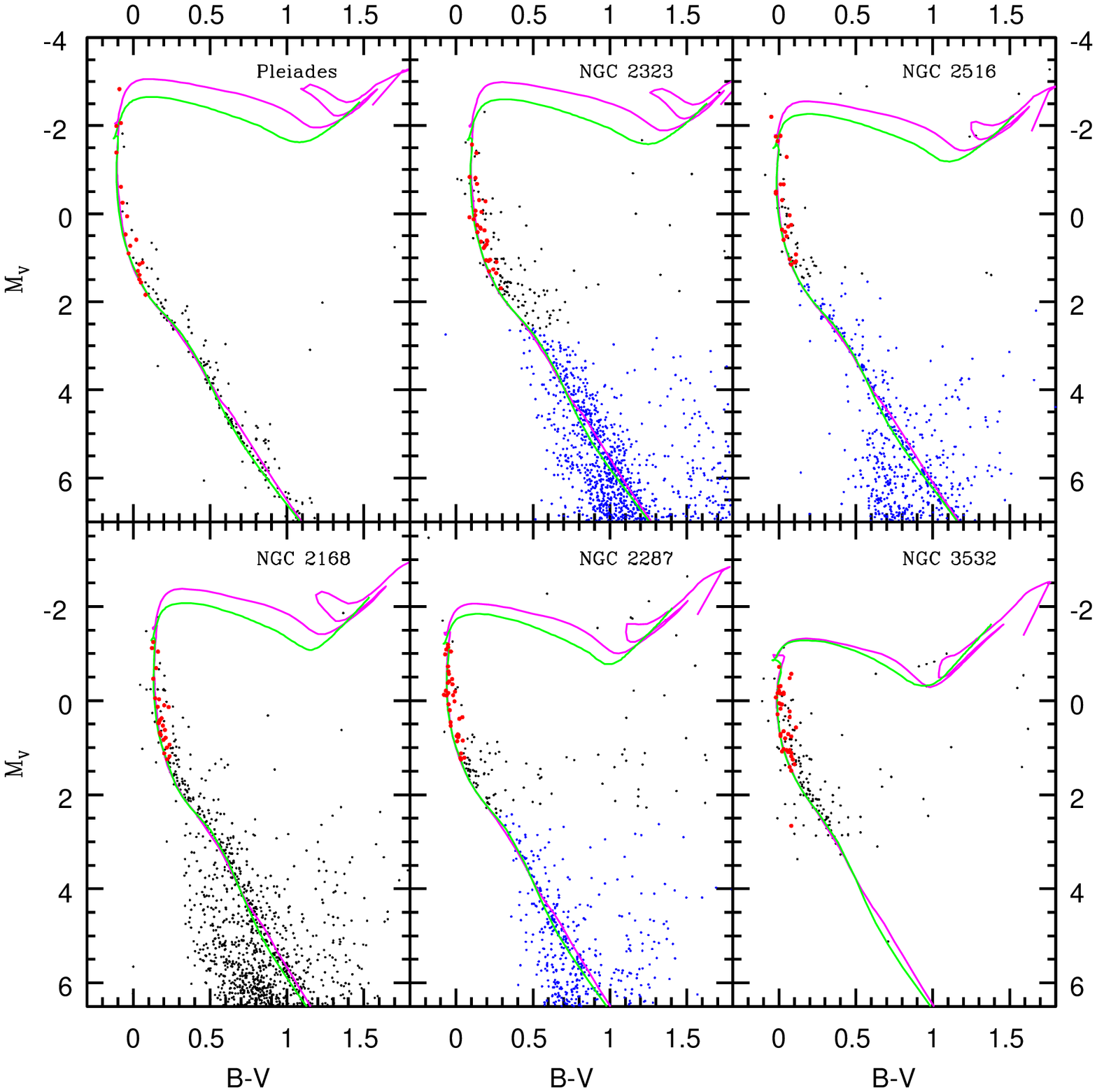}
\end{center}
\vspace{-0.4cm}
\caption{Our color-magnitude diagrams for our six clusters.  The full data sets are
shown in black, with several clusters having supplemental data shown in blue.  Our final 
turnoff stars selected in Figure 3 are again shown here in red, and we show our Y$^2$
isochrone fits in green and our PARSEC isochrone fits in magenta.  See Table 2 and
the text for our parameters and photometric sources.}
\end{figure*}

\tablefontsize{\footnotesize}
\begin{center}
\begin{deluxetable*}{l c c c c c c}
\multicolumn{7}{c}%
{{\bfseries \tablename\ \thetable{} - Open Cluster Parameters}} \\
\hline
Cluster& E(B-V)$^a$ & (m-M)$_0$& [Fe/H] & Y$^2$ Age (Myr) & PARSEC Age (Myr) & Photometric Sources\\
\hline
Pleiades & 0.03$\pm$0.02   & 5.67$\pm$0.10  & +0.01   & 135$\pm$15  & 110$\pm$15 & 1\\
NGC 2323 & 0.23$\pm$0.06   & 10.0$\pm$0.15  & 0.00    & 140$\pm$20  & 115$\pm$20 & 2,3\\
NGC 2516 & 0.10$\pm$0.03   & 8.20$\pm$0.12  & +0.065  & 170$\pm$20  & 150$\pm$20 & 4,5\\
NGC 2168 & 0.25$\pm$0.04   & 9.66$\pm$0.10  & -0.143  & 190$\pm$20  & 170$\pm$20 & 6\\
NGC 2287 & 0.035$\pm$0.025 & 9.52$\pm$0.12  & 0.00    & 220$\pm$30  & 205$\pm$30 & 7,8\\
NGC 3532 & 0.04$\pm$0.025  & 8.46$\pm$0.14  & 0.00    & 320$\pm$20  & 325$\pm$20 & 9\\
\hline
\caption{a) For reddenings of 0.10 or larger we have adopted the color dependent reddening relation
of Fernie (1963) and give the derived reddenings at a color of (B-V)$_0$=0.  If there are more than 
two photometric sources, the primary source is listed first and the secondary source is only for 
faint stars beyond the photometric limit
of the primary. (1) Johnson \& Mitchell (1958); (2) Claria et~al.\ (1998); (3) Kalirai et~al.\ (2003);
(4) Dachs (1970); (5) Sung et~al.\ (2002); (6) Sung \& Bessell (1999); (7) Ianna et~al.\ (1987);
(8) Sharma et~al.\ (2006); (9) Fernandez \& Salgado (1980).}
\end{deluxetable*}
\end{center}

\vspace{-1.9cm}

\tablefontsize{\footnotesize}
\begin{center}
\begin{deluxetable*}{l c c c c c c c c}
\multicolumn{9}{c}%
{{\bfseries \tablename\ \thetable{} - Membership Data}} \\
\hline
ID&$\alpha$&$\delta$&M$_V$       &V     &(B-V)$_0$   &B-V   &t$_{cool}$\\
  &(J2000) &(J2000) &(Model)&(Obs.)&(Model)&(Obs.)&(Myr)\\
\hline
\multicolumn{9}{l}{{NGC 2323 Likely Single Star White Dwarf Members}}  \\
\hline
NGC 2323-WD10  & 7:02:41.02 & -8:26:12.8  & 10.36$\pm$0.19 & 20.62$\pm$0.009 &  -0.303$\pm$0.002  &  -0.020$\pm$0.060  & 2.9$^{+1.9}_{-1.0}$\\
NGC 2323-WD11  & 7:03:22.14 & -8:15:58.7  & 10.36$\pm$0.13 & 20.67$\pm$0.008 &  -0.305$\pm$0.002  &   0.050$\pm$0.025  & 2.6$^{+1.1}_{-0.7}$\\
\hline
\multicolumn{9}{l}{{White Dwarfs Inconsistent with Single Star Membership}}  \\
\hline
NGC 2323-WD21  & 7:02:08.56 & -8:25:48.0  & 11.33$\pm$0.25 & 21.91$\pm$0.026 &  -0.013$\pm$0.022  &   0.149$\pm$0.158  & 170$^{+60}_{-48}$\\ 
NGC 2323-WD7   & 7:02:47.87 & -8:35:56.4  & 10.92$\pm$0.07 & 19.50$\pm$0.004 &  -0.005$\pm$0.007  &   0.078$\pm$0.004  & 112$^{+12}_{-11}$\\ 
NGC 2323-WD17  & 7:03:12.62 & -8:30:53.6  & 10.97$\pm$0.07 & 20.96$\pm$0.010 &  -0.057$\pm$0.007  &   0.107$\pm$0.015  & 96$^{+12}_{-11}$\\  
NGC 2323-WD12  & 7:03:17.79 & -8:19:59.7  & 10.87$\pm$0.11 & 20.52$\pm$0.007 &  -0.014$\pm$0.010  &   0.116$\pm$0.079  & 101$^{+17}_{-15}$\\ 
\hline
\multicolumn{9}{l}{{Low Signal to Noise and Featureless White Dwarfs}}  \\
\hline
NGC 2323-WD22  & 7:02:33.93 & -8:31:04.3  & 10.53$\pm$0.36 & 21.89$\pm$0.024 &  -0.135$\pm$0.025  &   0.180$\pm$0.013  & 33$^{+36}_{-16}$\\  
NGC 2323-WD30  & 7:03:22.56 & -8:29:25.7  & 11.52$\pm$0.28 & 22.04$\pm$0.028 &   0.111$\pm$0.026  &   0.485$\pm$0.029  & 290$^{+108}_{-82}$\\
NGC 2323-WD38  & 7:03:31.87 & -8:28:25.0  &  --            & 22.93$\pm$0.062 &        --          &   0.645$\pm$0.058  &    --  \\ 
NGC 2323-WD23  & 7:03:39.85 & -8:28:16.7  &  --            & 21.42$\pm$0.016 &        --          &   0.375$\pm$0.044  &    --  \\ 
\hline
\end{deluxetable*}
\end{center}

Our four additional clusters are also shown in Figure 3.  In NGC 2516, we have used two 
UBV photometric studies, Dachs (1970) for the brightest stars and Sung et~al.\ (2002)
for the fainter stars shown in blue.  For the moderately large reddening we fit a E(B-V) 
of 0.10$\pm$0.03 at (B-V)$_0$=0 and a [Fe/H]=0.065, and as with the Pleiades this is consistent
with the typically adopted parameters and our spectroscopic analysis (Cummings 2011).  
For NGC 2287 we have used the UBV photometry of Ianna et~al.\ (1987) and fit a E(B-V) of 
0.035$\pm$0.025 with an assumed solar metallicity.  For NGC 3532 we have used the UBV photometry 
of Fernandez \& Salgado (1980) and fit a E(B-V) of 0.04$\pm$0.025 with an assumed solar metallicity.  
Lastly, for NGC 2168 we have used the UBV photometry of Sung \& Bessell (1999).  We derive a E(B-V) 
of 0.25$\pm$0.04 at (B-V)$_0$=0 for this cluster and adopt a metal-poor [Fe/H] of -0.143 (Steinhauer 
\& Deliyannis 2004), which provides a reliable fit to the bluest stars and follows the
U-B range well but it appears that there is a systematic shift in B-V where the stars
becoming increasingly too blue at redder colors.  Otherwise, changes in either adopted 
metallicity or reddening cannot fit the full B-V color range.

\vspace{0.2cm}
\subsection{Color-Magnitude Analysis}

In Figure 4 we display our by eye turnoff age fits with Y$^2$ isochrones in green 
when adopting from Figure 3 the reddenings, metallicities, and
the cleaned turnoff stars shown in red.  In magenta we similarly display the PARSEC isochrone
fits, where we have adopted the same reddenings, distance moduli, and metallicities (Z).  In Table 2 
the derived cluster parameters and the photometric sources are listed.  
We note that in the youngest clusters the PARSEC isochrones systematically derive ages
25 Myr younger, while in the 320 Myr NGC 3532 they derive ages 5 Myr older, and lastly from Paper I
the PARSEC isochrones derive an age 20 Myr older in the 520 Myr old NGC 2099.  Therefore, these
two isochrones not only have changing systematics at differing ages, but they are in opposite
directions in young versus older clusters.  The possible causes of the systematics between these
two isochrones include the differences in their adopted opacities, equations of state, and solar 
compositions.  While the Y$^2$ isochrones do not consider evolution past the tip of the red giant
branch (RGB), in these typically young clusters we cannot reliably fit the giants because their 
populations are very sparse or they have no giants at all.  Therefore, we have chosen the Y$^2$ 
isochrones for our final parameters because they more successfully fit both the color-color data and the
the main sequence features. 

These cluster ages provide several advantages over adopting literature values.  1)
The ages are based on a uniform system of isochrones, while literature values adopt wide ranging
models that have systematic differences that become more pronounced at younger ages.  2) The
fitting techniques applied are by eye but consistent, while fitting techniques for literature values 
can greatly vary.  3) The difficulty of peculiar
turnoff stars are addressed in a systematic way, while their consideration can have important differences
in the literature values, if they are considered at all.  We will not comment on the absolute accuracy of the 
various isochrone model ages, but in this study uniformity and precision is the goal.  We must also
reiterate two remaining limitations with our cluster parameters.  First, all of our
photometries are primarily from differing groups, which still may leave important systematics
remaining in our parameter analyses.  Additionally, uniformly measured spectroscopic metallicities
are also needed to address the metallicity sensitivity in the turnoff isochrone fits.

Lastly, for Sirius B there is no parent cluster that we can self-consistently analyze for
the total age, but the age of the Sirius system is well studied and here we adopt solar metallicity
and the age of 237.5$\pm$12.5 Myr determined in Liebert et~al.\ (2005).

\section{White Dwarf Membership in NGC 2323}

\begin{figure*}[!ht]
\begin{center}
\includegraphics[clip, scale=0.9]{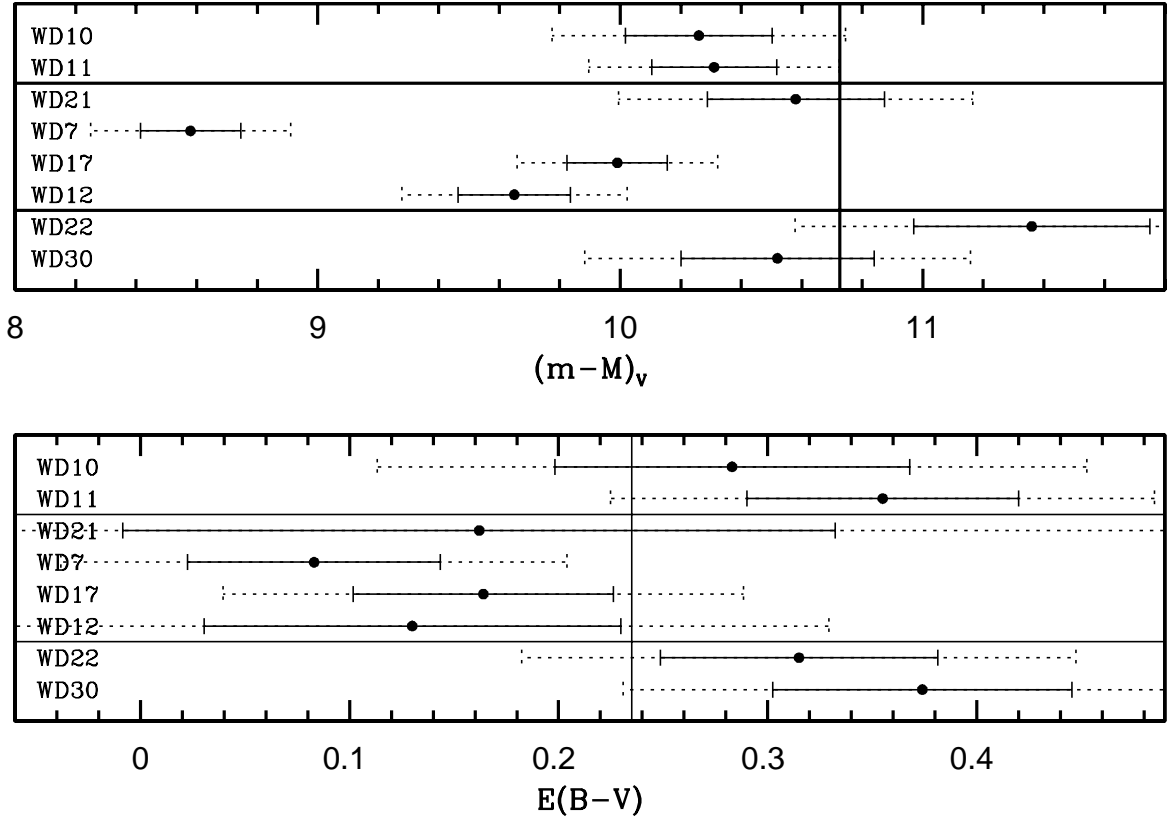}
\end{center}
\vspace{-0.4cm}
\caption{The upper panel compares the model based and observed white dwarf magnitudes relative 
to the cluster distance modulus of 10.725 (adjusted based on their blue colors).  Similarly, 
the lower panel compares the model based and observed white dwarf B-V colors relative to the 
cluster reddening of 0.235 (adjusted based on the blue colors).  In both panels the respective
1$\sigma$ error bars are shown, which include both the spectroscopic and photometric errors.  
WD10, WD11, and WD21 are shown at the top and are consistent (within 2$\sigma$) with cluster 
membership in both magnitude and color, but WD21 has too long of a cooling age for NGC 2323
and has been grouped with the three additional nonmembers displayed in the middle.  Lastly, 
the bottom two white dwarfs are displayed but have spectroscopic fitting errors above our cut 
for membership analysis.}
\end{figure*}

\begin{figure}[!ht]
\begin{center}
\includegraphics[clip, scale=0.87]{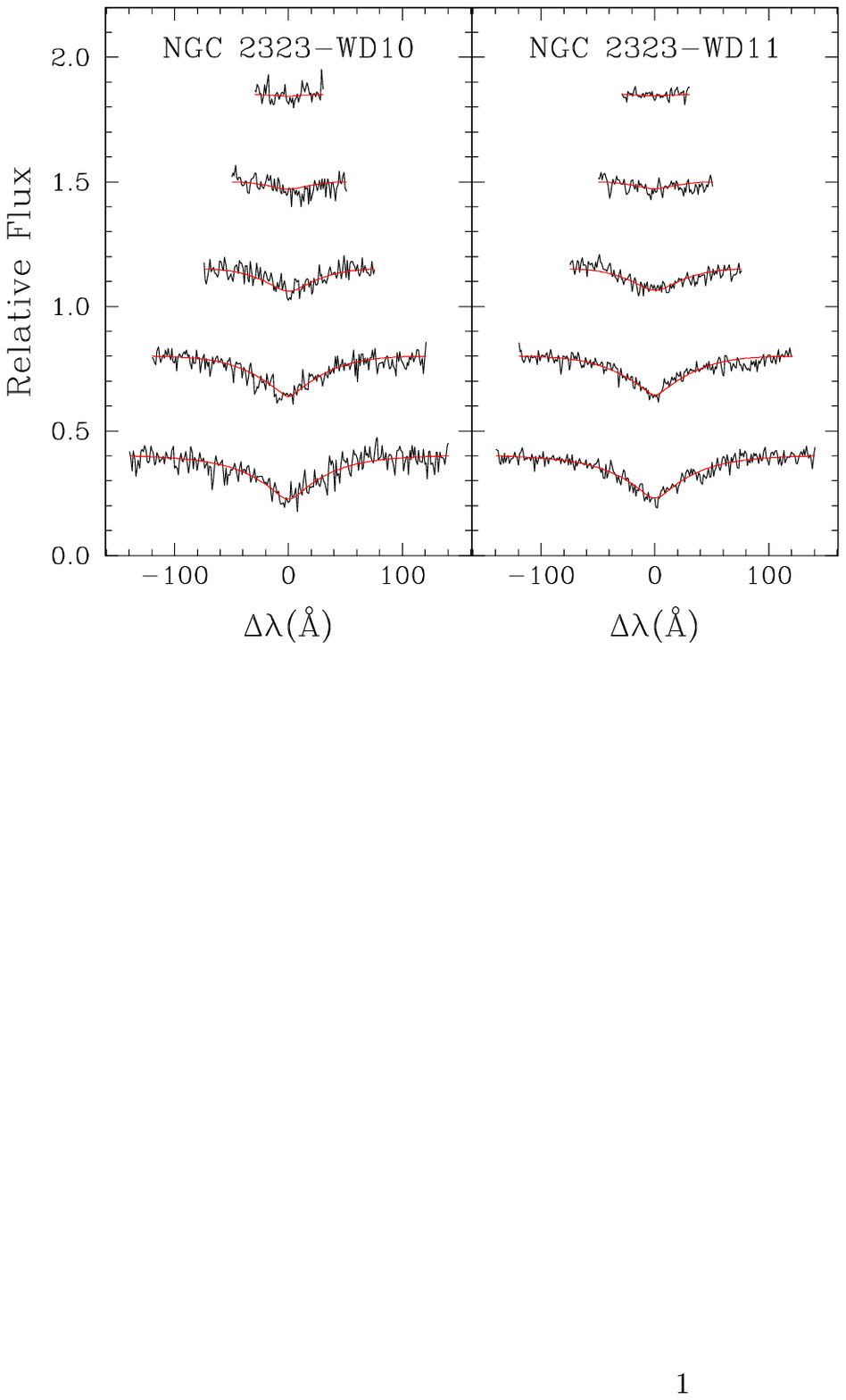}
\end{center}
\vspace{-0.4cm}
\caption{The Balmer line fits for the two white dwarf members of NGC 2323.  The H$\beta$, 
H$\gamma$, H$\delta$, H$\epsilon$, and H8 fits are shown from bottom to top.}
\end{figure}

\begin{figure}[!ht]
\begin{center}
\includegraphics[scale=0.435]{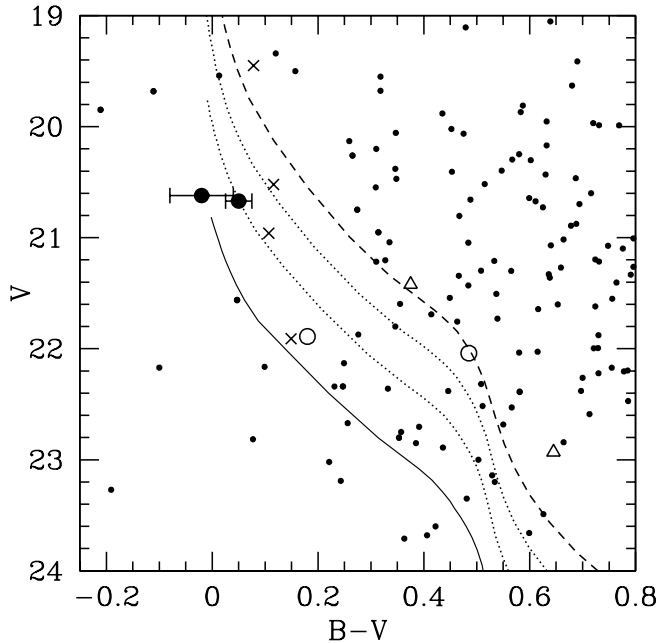}
\end{center}
\vspace{-0.4cm}
\caption{The white dwarf photometry where we differentiate between observed non-members 
(x's), likely singly evolved members (solid circle), observed white dwarfs with low S/N 
(open circles), featureless spectra (triangles), and unobserved candidates (small points).  
White dwarf cooling models are plotted for reference from left to right of 1.2 M$_\odot$ 
(solid line), 1.0, 0.8, and 0.6 M$_\odot$ (dotted lines), and 0.4 M$_\odot$ (dashed line).  
The mean distance modulus and reddening derived from the white dwarf members are applied 
to these cooling models.}
\end{figure}

\begin{figure*}[!ht]
\begin{center}
\includegraphics[clip, scale=0.9]{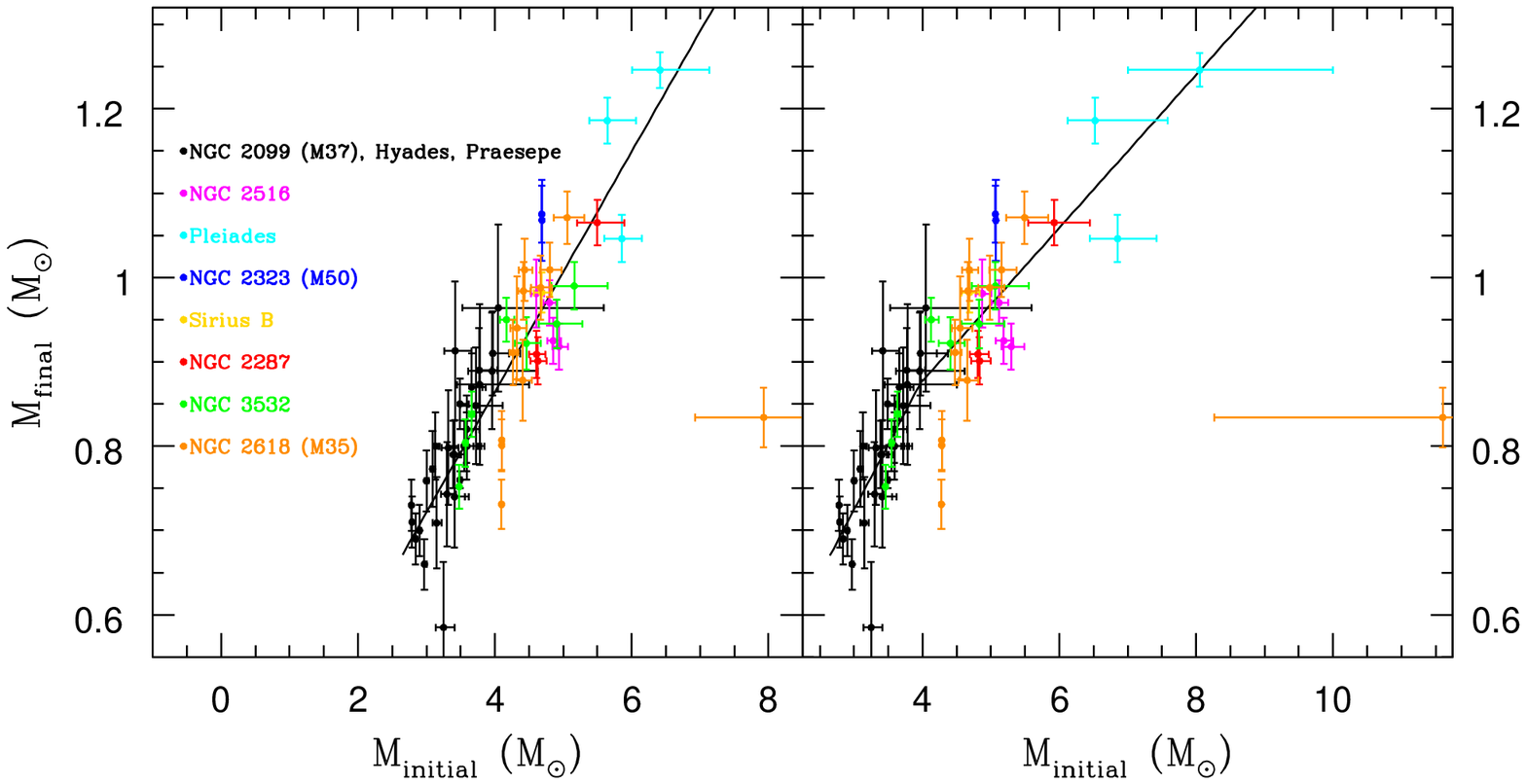}
\end{center}
\vspace{-0.4cm}
\caption{The left panel shows our IFMR for our analysis based on the Y$^2$ cluster ages,
with our studied clusters each color coded and compared to our results from Paper I.  The 
weighted linear fit is shown in black.  The right panel shows our IFMR for our PARSEC cluster 
ages, which are typically younger and give higher-mass progenitors.
This gives a clear turnover in the IFMR at higher masses, which we fit using a two-piece linear
function shown in black.  As discussed in the text, the results from Paper I are determined 
using the same methods and with cluster ages derived from Y$^2$ isochrones, but at these intermediate
masses the difference between the Y$^2$ isochrones and PARSEC isochrones are not significant.}
\end{figure*}

Cluster membership determination is key in analyzing the formation history of these white 
dwarfs and applying them to the IFMR.  In the case of NGC 2323, we have the advantage that 
because it is such a young cluster, any hot and young high-mass white dwarfs observed in 
its field would already have reliable membership.  For example, in the Sloan Digital Sky
Survey sample of field white dwarfs, only $\sim$2.6\% of DA white dwarfs 
have a mass greater than 1 M$_\odot$ and less than 10\% of these are young with a 
T$_{\rm eff}$$>$20,000 K (e.g., Kleinman et~al.\ 2013; Kepler et~al.\ 2016).  To 
further confirm cluster membership in NGC 2323, however, we have also performed magnitude and color 
analysis.

In Table 3 we list the observed photometry for these white dwarfs and also list their 
cooling age and model based M$_V$ and (B-V)$_0$.  For our membership analysis we 
directly compare the model based and observed magnitudes and colors in Figure 5 to 
derive an effective distance modulus and reddening for each white dwarf.  We define 
our 1$\sigma$ color and magnitude errors by adding in quadrature the respective model 
fitting errors, the observational errors, and in the case of magnitude the distance 
modulus errors or in the case of color the reddening errors.  We select white dwarfs 
as consistent with membership if their effective distance modulus and reddening are 
within 2$\sigma$ of the cluster parameters derived in Section 3.  
Of our six high signal white dwarfs, only WD10, WD11, and WD21 have both magnitudes and
colors consistent with membership.  However, WD21 has a cooling age longer than the age
of the cluster.  Therefore, we do not consider it a member but a field white dwarf 
at comparable distance to NGC 2323.

In Figure 6 we show the Balmer line fits of WD10 and WD11, the two cluster members.  They 
both are high mass at $\sim$1.07 M$_\odot$ and have very high temperature ($>$50,000 K) 
with accordingly extremely short cooling times of $\sim$3 Myr.  Their membership further 
suggests that all of our other observed white dwarfs, which all have both lower masses and longer 
cooling times are not consistent with NGC 2323 membership.  In Figure 7 we display the CMD of 
all observed candidates relative to several white dwarf cooling models taken from 
http://www.astro.umontreal.ca/{\raise.17ex\hbox{$\scriptstyle\sim$}}bergeron/CoolingModels/ (Holberg \& Bergeron 2006, Kowalski \& 
Saumon 2006, Tremblay et~al.\ 2011, and Bergeron et~al.\ 2011).

\section{Initial-Final Mass Relation}

\begin{center}
\begin{deluxetable*}{l c c c c c c c c}
\multicolumn{9}{c}%
{{\bfseries \tablename\ \thetable{} - Reanalyzed White Dwarf Initial and Final Parameters}} \\
\hline
ID&T$_{\rm eff}$&log g&M$_{WD}$   & M$_V$ &t$_{cool}$& Y$^2$ M$_i$ & PARSEC M$_i$ & S/N\\
  &(K)          &     &(M$_\odot$)&       &(Myr)     &(M$_\odot$)  &(M$_\odot$)   &\\
\hline
NGC 2287-2         & 25900$\pm$350  & 8.45$\pm$0.05 & 0.909$\pm$0.028 & 11.01$\pm$0.09 &  76$^{+10}_{-9}$    & 4.61$^{+0.13}_{-0.11}$ & 4.81$^{+0.16}_{-0.12}$ & 150\\
NGC 2287-4         & 26500$\pm$350  & 8.71$\pm$0.05 & 1.065$\pm$0.027 & 11.44$\pm$0.10 & 127$^{+14}_{-13}$   & 5.50$^{+0.40}_{-0.30}$ & 5.93$^{+0.52}_{-0.38}$ & 130\\
NGC 2287-5         & 25600$\pm$350  & 8.44$\pm$0.04 & 0.901$\pm$0.028 & 11.02$\pm$0.08 &  77$^{+10}_{-9}$    & 4.63$^{+0.13}_{-0.11}$ & 4.83$^{+0.16}_{-0.12}$ & 170\\
NGC 2516-1         & 30100$\pm$350  & 8.47$\pm$0.04 & 0.925$\pm$0.027 & 10.74$\pm$0.08 &  42$^{+7}_{-7}$     & 4.85$^{+0.11}_{-0.09}$ & 5.19$^{+0.14}_{-0.12}$ & 160\\
NGC 2516-2         & 35500$\pm$550  & 8.55$\pm$0.07 & 0.981$\pm$0.040 & 10.58$\pm$0.13 &  24$^{+8}_{-7}$     & 4.61$^{+0.11}_{-0.08}$ & 4.88$^{+0.13}_{-0.10}$ &  90\\
NGC 2516-3         & 29100$\pm$350  & 8.46$\pm$0.04 & 0.918$\pm$0.027 & 10.78$\pm$0.08 &  48$^{+8}_{-7}$     & 4.94$^{+0.13}_{-0.10}$ & 5.31$^{+0.19}_{-0.14}$ & 170\\
NGC 2516-5         & 32200$\pm$400  & 8.54$\pm$0.05 & 0.970$\pm$0.027 & 10.73$\pm$0.09 &  38$^{+7}_{-6}$     & 4.80$^{+0.10}_{-0.09}$ & 5.12$^{+0.13}_{-0.11}$ & 170\\
NGC 3532-1         & 23100$\pm$300  & 8.52$\pm$0.04 & 0.950$\pm$0.026 & 11.33$\pm$0.08 & 131$^{+13}_{-12}$   & 4.17$^{+0.11}_{-0.09}$ & 4.13$^{+0.11}_{-0.09}$ & 240\\
NGC 3532-5         & 27700$\pm$350  & 8.28$\pm$0.05 & 0.804$\pm$0.028 & 10.57$\pm$0.08 &  31$^{+7}_{-6}$     & 3.57$^{+0.03}_{-0.03}$ & 3.55$^{+0.03}_{-0.03}$ & 220\\
NGC 3532-9         & 31900$\pm$400  & 8.18$\pm$0.04 & 0.752$\pm$0.026 & 10.13$\pm$0.07 & 9.3$^{+2}_{-1}$     & 3.48$^{+0.01}_{-0.01}$ & 3.46$^{+0.01}_{-0.01}$ & 210\\
NGC 3532-10        & 26300$\pm$350  & 8.34$\pm$0.04 & 0.838$\pm$0.027 & 10.78$\pm$0.08 &  51$^{+8}_{-8}$     & 3.66$^{+0.04}_{-0.04}$ & 3.64$^{+0.04}_{-0.04}$ & 200\\
NGC 3532-J1106-590 & 21100$\pm$350  & 8.48$\pm$0.05 & 0.922$\pm$0.031 & 11.43$\pm$0.09 & 163$^{+18}_{-17}$   & 4.46$^{+0.21}_{-0.17}$ & 4.41$^{+0.21}_{-0.17}$ & 110\\
NGC 3532-J1106-584 & 20200$\pm$300  & 8.52$\pm$0.05 & 0.945$\pm$0.029 & 11.58$\pm$0.09 & 197$^{+20}_{-18}$   & 4.91$^{+0.37}_{-0.26}$ & 4.83$^{+0.37}_{-0.26}$ & 120\\
NGC 3532-J1107-584 & 20700$\pm$300  & 8.59$\pm$0.05 & 0.990$\pm$0.028 & 11.66$\pm$0.09 & 211$^{+21}_{-20}$   & 5.16$^{+0.49}_{-0.34}$ & 5.07$^{+0.49}_{-0.34}$ & 180\\
Sirius B           & 26000$\pm$400  & 8.57$\pm$0.04 & 0.982$\pm$0.024 & 11.21$\pm$0.08 &  99$^{+11}_{-10}$   & 4.69$^{+0.15}_{-0.12}$ & 4.69$^{+0.15}_{-0.12}$ &  - \\
Pleiades-LB 1497   & 32700$\pm$500  & 8.67$\pm$0.05 & 1.046$\pm$0.028 & 10.94$\pm$0.09 &  54$^{+9}_{-8}$     & 5.86$^{+0.29}_{-0.26}$ & 6.85$^{+0.57}_{-0.41}$ & 170\\
Pleiades-GD50      & 42700$\pm$800  & 9.20$\pm$0.07 & 1.246$\pm$0.021 & 11.58$\pm$0.15 &  70$^{+14}_{-13}$   & 6.41$^{+0.72}_{-0.41}$ & 8.05$^{+1.95}_{-1.05}$ &  - \\
Pleiades-PG0136+251& 41400$\pm$800  & 9.03$\pm$0.07 & 1.186$\pm$0.027 & 11.28$\pm$0.15 &  48$^{+12}_{-11}$   & 5.64$^{+0.42}_{-0.26}$ & 6.52$^{+1.06}_{-0.40}$ &  - \\
NGC 2168-LAWDS1    & 33500$\pm$450  & 8.44$\pm$0.06 & 0.911$\pm$0.039 & 10.47$\pm$0.11 &  19$^{+7}_{-6}$     & 4.27$^{+0.07}_{-0.05}$ & 4.48$^{+0.09}_{-0.06}$ &  95\\
NGC 2168-LAWDS2    & 33400$\pm$600  & 8.49$\pm$0.10 & 0.940$\pm$0.061 & 10.57$\pm$0.18 &  25$^{+13}_{-10}$   & 4.33$^{+0.14}_{-0.10}$ & 4.55$^{+0.18}_{-0.12}$ &  70\\
NGC 2168-LAWDS5    & 52700$\pm$900  & 8.21$\pm$0.06 & 0.801$\pm$0.031 &  9.49$\pm$0.10 & 1.0$^{+0.1}_{-0.1}$ & 4.10$^{+0.01}_{-0.01}$ & 4.28$^{+0.01}_{-0.01}$ & 210\\
NGC 2168-LAWDS6    & 57300$\pm$1000 & 8.05$\pm$0.06 & 0.731$\pm$0.029 &  9.13$\pm$0.11 & 0.5$^{+0.1}_{-0.1}$ & 4.10$^{+0.01}_{-0.01}$ & 4.28$^{+0.01}_{-0.01}$ & 260\\
NGC 2168-LAWDS11   & 19900$\pm$350  & 8.35$\pm$0.05 & 0.834$\pm$0.035 & 11.31$\pm$0.09 & 149$^{+18}_{-17}$   & 7.93$^{+1.44}_{-1.00}$ &11.60$^{+*}_{-3.34}$    & 100\\
NGC 2168-LAWDS12   & 34200$\pm$500  & 8.60$\pm$0.06 & 1.009$\pm$0.037 & 10.73$\pm$0.12 &  36$^{+9}_{-8}$     & 4.44$^{+0.11}_{-0.09}$ & 4.69$^{+0.13}_{-0.11}$ & 100\\
NGC 2168-LAWDS14   & 30500$\pm$450  & 8.57$\pm$0.06 & 0.988$\pm$0.038 & 10.89$\pm$0.12 &  54$^{+11}_{-10}$   & 4.67$^{+0.16}_{-0.14}$ & 4.98$^{+0.21}_{-0.17}$ & 100\\
NGC 2168-LAWDS15   & 30100$\pm$400  & 8.61$\pm$0.06 & 1.009$\pm$0.032 & 10.98$\pm$0.11 &  64$^{+10}_{-10}$   & 4.80$^{+0.17}_{-0.14}$ & 5.16$^{+0.22}_{-0.18}$ & 130\\
NGC 2168-LAWDS22   & 53000$\pm$1000 & 8.22$\pm$0.06 & 0.807$\pm$0.035 &  9.50$\pm$0.11 & 1.0$^{+0.1}_{-0.1}$ & 4.10$^{+0.01}_{-0.01}$ & 4.28$^{+0.01}_{-0.01}$ & 250\\
NGC 2168-LAWDS27   & 30700$\pm$400  & 8.72$\pm$0.06 & 1.071$\pm$0.031 & 11.16$\pm$0.11 &  78$^{+12}_{-11}$   & 5.06$^{+0.25}_{-0.20}$ & 5.49$^{+0.35}_{-0.27}$ & 120\\
NGC 2168-LAWDS29   & 33500$\pm$450  & 8.56$\pm$0.06 & 0.984$\pm$0.034 & 10.70$\pm$0.11 &  34$^{+8}_{-8}$     & 4.42$^{+0.10}_{-0.08}$ & 4.67$^{+0.12}_{-0.10}$ & 120\\
NGC 2168-LAWDS30   & 29700$\pm$500  & 8.39$\pm$0.08 & 0.878$\pm$0.048 & 10.63$\pm$0.13 &  33$^{+12}_{-10}$   & 4.41$^{+0.13}_{-0.10}$ & 4.65$^{+0.16}_{-0.12}$ &  60\\
\hline
\end{deluxetable*}
\end{center}

\vspace{-0.8cm}
With the white dwarf cluster members, a simple comparison of their cooling age to the total 
cluster age gives the evolutionary time to the tip of the asymptotic giant branch (AGB) for their
progenitor.  Application of this time to evolutionary models gives the white dwarf's progenitor 
mass.  In Paper I we adopted the models of Hurley et~al.\ (2000) for the progenitor 
masses in NGC 2099 (3-4 M$_\odot$).  At these masses, the difference between predicted progenitor
mass by the Hurley et~al.\ (2000) models and the PARSEC isochrones are less than 1\%.  We
note that the PARSEC models do not include the thermally pulsing-AGB phase, but this phase
is very short and does not meaningfully affect the resulting progenitor masses.  For the Y$^2$ 
isochrones we cannot infer progenitor masses directly because these isochrones
do not evolve beyond the RGB.  But we note that while the Y$^2$ isochrones 
predict slower evolution to the turnoff in these younger clusters, they predict more 
rapid evolution through the RGB.  This results in the total evolutionary time 
scales to the tip of the RGB being comparable in both model isochrones for
all but the highest masses.  For example, as we reach evolutionary times of 100 Myr 
or shorter (progenitors of $\gtrsim$5.3 M$_\odot$), the systematic differences become 
significant between all three models.  For our current analysis, with both our
Y$^2$ and PARSEC ages, we use progenitor masses derived from the PARSEC isochrones 
because it will provide the strongest consistency with our cluster age fits.

For NGC 2323-WD10 and WD11, with their cooling times and our Y$^2$ isochrones based age 
for NGC 2323 of 140 Myr, the corresponding progenitor masses for both white dwarfs are 
4.69 M$_\odot$. (See Table 1, where we also give the progenitor masses based on the cluster 
age errors [140$\pm$20 Myr] and the PARSEC based age of 115 Myr.) 
It is quite remarkable that we have two independently formed high-mass white dwarfs from the same 
cluster that are so consistent in both initial and final mass.  Across their lifetimes they both lost 
77.2\% of their total mass.  This argues for the consistency of single-star 
mass loss at higher masses.  To look at these data in context, in Table 4 we present the 
initial-final mass data for 30 mid to very high-mass (0.73-1.25 M$_\odot$)
white dwarfs that have been self-consistently analyzed from publicly available data.  

In the left panel of Figure 8 we plot the initial and final masses of the two analyzed white dwarfs
from NGC 2323 and the 30 white dwarfs we have reanalyzed from the literature, adopting the
Y$^2$ ages.  In the right panel of Figure 8 we plot these same data, but with application of
the PARSEC based ages.  We also compare these high-mass white dwarf data to the sample of 
31 intermediate-mass white dwarfs taken from NGC 2099, the Hyades, and Praesepe (Paper I).
For the higher-mass white dwarfs, while there is some dispersion in the trend there is
only one clear outlier that is from NGC 2168.  LAWDS11 is an extreme outlier with a far longer
cooling time than the other members of NGC 2168, giving it a massive progenitor.  For clusters 
in the rich galactic plane, contamination from common field white dwarfs is
expected and likely explains this white dwarf.  Proper motions may be necessary to further constrain
its membership, but in our current analysis and IFMR fits we do not consider LAWDS11.

In our initial analysis from Paper I, we demonstrated that the intermediate-mass
white dwarfs (0.7--0.9 M$_\odot$) create a steep IFMR slope.  With the continued adoption of 
Y$^2$ ages in this paper, which systematically derive older ages in young clusters, 
we find there is no meaningful change in slope at higher 
masses and the data appears to be defined well with a weighted linear relation:
\vspace{-0.1cm}
\begin{equation*} 
M_{\rm final}=(0.143\pm0.005)M_{\rm initial}+0.294\pm0.020 M_\odot.
\vspace{-0.05cm}
\end{equation*}
The linear nature of the IFMR across such a broad range (3--6.5 M$_{\rm initial}$) is of interest.
For example, this may suggest we can extrapolate to derive the progenitor of a Chandrasekhar
mass limit white dwarf to be $\sim$7.75 M$_\odot$, but the still limited data at the highest masses 
and the remaining uncertainties in the evolutionary models suggests this is unreliable.  Furthermore, in 
theoretical models a moderate turnover in the slope of the IFMR is predicted near initial mass 
of 4 M$_\odot$ (e.g., Marigo \& Girardi 2007 and Meng et~al.\ 2008).  This predicted turnover is the result 
of the second dredge-up, which only occurs in stars of $\sim$4 M$_\odot$ and higher.  This diminishes 
their core mass and hence their final white dwarf mass.  A comparable turnover could still be lost
in our data's remaining scatter, causing further problems with a linear extrapolation.  In Figure 9 we
more closely analyze the residuals of this linear data fit.

\begin{figure}[!ht]
\begin{center}
\includegraphics[scale=0.435]{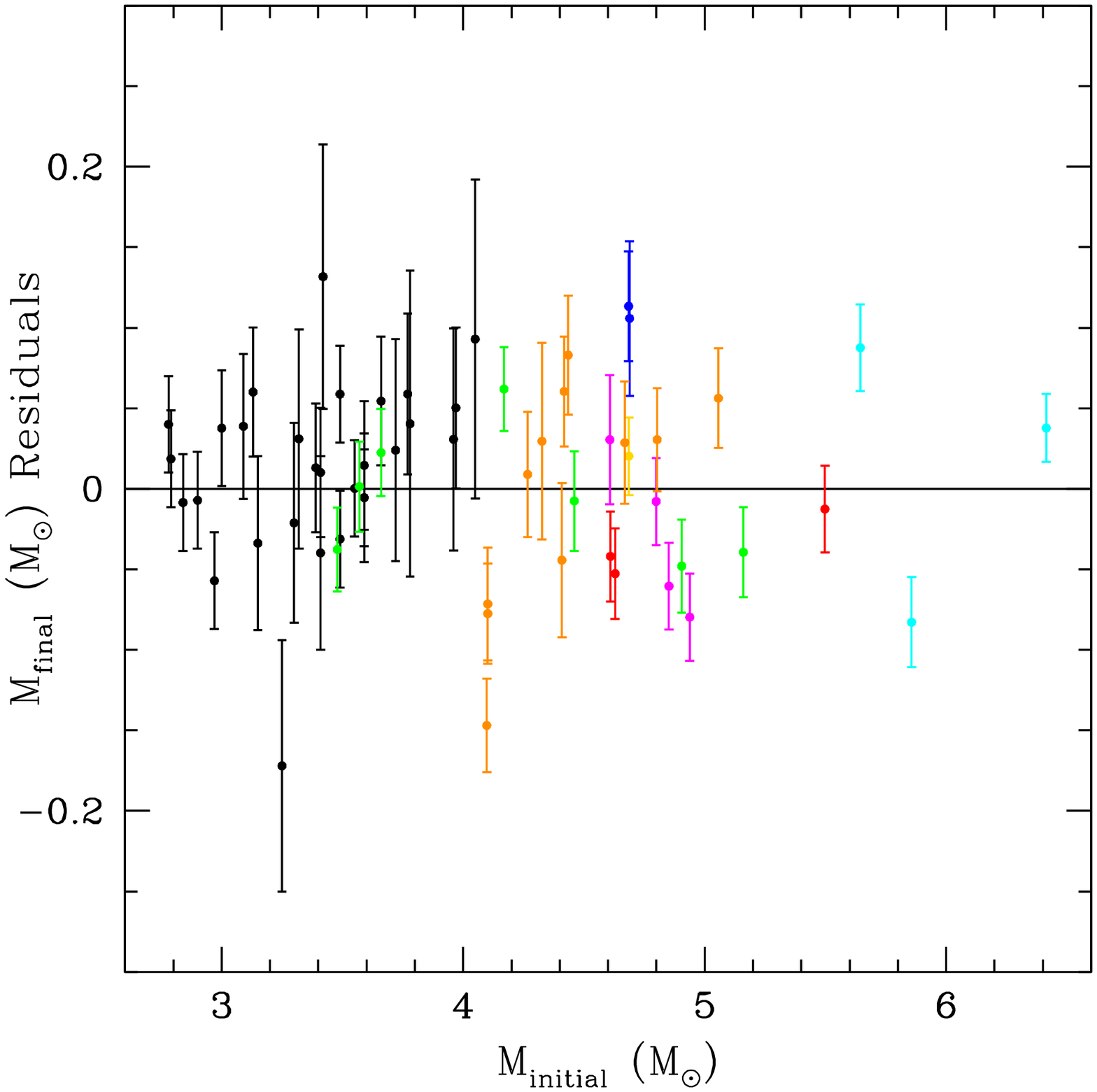}
\end{center}
\vspace{-0.4cm}
\caption{The residuals in white dwarf masses are plotted against initial mass.  The same color
scheme from Figure 8 is adopted.  The residuals are normally distributed and show that when considering 
the errors and the remaining scatter, the simple linear fit represents the data well.}
\end{figure}

To illustrate the importance of our adopted isochrones, the right panel of Figure 8 shows a clear 
turnover in the IFMR near an initial mass of 4 M$_\odot$ when adopting PARSEC isochrone based ages,
consistent with theoretical predictions.  We note that while our comparison IFMR data from Paper I 
adopts Y$^2$ isochrone ages, this is not the 
cause of the strong overturn.  This is because in these older clusters ages from Paper I 
the systematics shift direction and the PARSEC isochrones give cluster ages 20 Myr older than 
the Y$^2$ isochrones.  For the intermediate-mass comparison IFMR this would increase its slope by 
$\sim$5\%, and this would further increase the magnitude of the turnover.  We fit the 
data based on the PARSEC ages for our younger clusters with a two-piece linear function with a 
kink at initial mass of 4 M$_\odot$:
\vspace{-0.3cm}
\begin{multline*} 
M_{\rm final}=(0.154\pm0.013)M_{\rm initial} \\
   +0.261\pm0.048 M_\odot (M_{\rm initial} < 4) \\ 
\hspace{-2.8cm} M_{\rm final}=(0.097\pm0.005)M_{\rm initial} \\
   +0.514\pm0.029 M_\odot (M_{\rm initial} \geq 4). \\ 
\end{multline*}
\vspace{-1cm}

\noindent This is a strong turnover, one that is stronger than predicted in Marigo \& Girardi (2007) 
or Meng et~al.\ (2008).  Taking a moment to focus specifically on NGC 3532 may help to 
clarify these systematic differences.  This is both because it is the only
cluster with progenitors that cover both above and below initial mass of 4 M$_\odot$,
and it is the cluster least effected by systematic differences between the Y$^2$ and PARSEC
isochrones.  The NGC 3532 data alone is limited but consistent with an overturn, which suggests
the Y$^2$ may be overestimating the ages of the youngest clusters, but the massive progenitors 
of the Pleiades white dwarfs in the right panel of Figure 8 conversely suggests that the PARSEC 
isochrones may be underestimating the ages of the older clusters.  Further work is needed on 
isochrones and evolutionary times at these higher masses, but for now we continue to adopt our 
Y$^2$ isochrone results as final.

\begin{figure}[!ht]
\begin{center}
\includegraphics[scale=0.435]{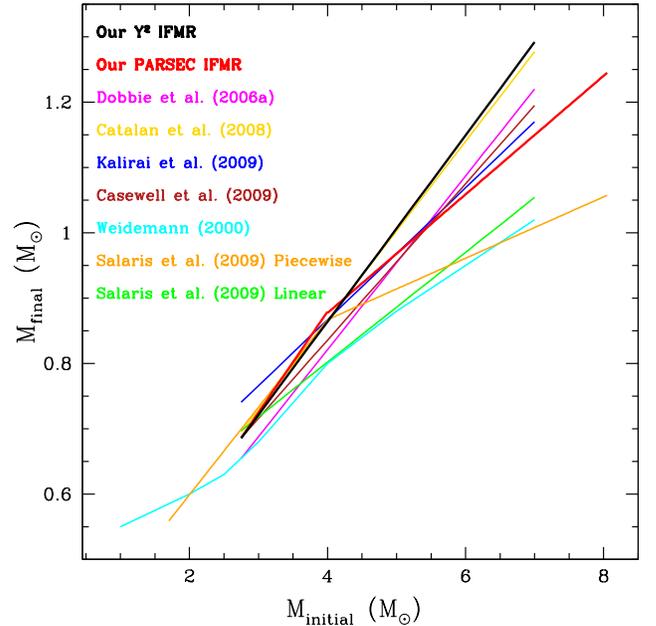}
\end{center}
\vspace{-0.4cm}
\caption{We compare our two IFMRs to the relations published in six other studies that are labeled in
the figure.}
\end{figure}

In Figure 10 we compare to multiple empirical IFMRs from Weidemann (2000), Dobbie et~al.\ (2006a), 
Catal{\'a}n et~al.\ (2008), Casewell et~al.\ (2009), Kalirai et~al.\ (2009), and Salaris et~al.\ (2009).
We find that our relation based on Y$^2$ ages gives that the IFMR is steeper 
at higher masses than all of these relations besides those of Dobbie et~al.\ (2006a) and Catal{\'a}n et~al.\
(2008), which have comparable slopes.  Our relation is strongly consistent with that of Catal{\'a}n 
et~al.\ (2008), but the zeropoint of Dobbie et~al.\ (2006a) is $\sim$0.05 M$_\odot$ lower.  Therefore, in 
comparison to most previous relations, ours predicts that white dwarf mass increases more rapidly with 
increasing progenitor mass, and that overall mass-loss rates are lower.  This finding is primarily the result 
of our ages being systematically older due in part to our analysis cleaning the turnoffs of peculiar
stars.  Additionally, the Y$^2$ isochrones give systematically older ages in comparison to the commonly 
adopted Girardi et~al.\ (2000) isochrones.  For example, a consistent turnoff fit to the Pleiades
with the isochrones of Girardi et~al. (2000) gives a 100 Myr age, 35 Myr younger than the Y$^2$ isochrones.  

The most remarkable difference to previous IFMRs is that the large scatter has significantly decreased  
by $\sim$50\%.  This not only illustrates the importance of doing uniform analysis of 
the white dwarf spectra, but of the consistent examination of the cluster parameters with detailed 
consideration of peculiar stars in the turnoffs.  Our IFMR's relatively smaller scatter also begins to 
show that while variations in mass-loss rates may still occur during a progenitor's lifetime, in singly 
evolved stars it is likely quite minor and the previously observed scatter is not indicative of stochastic
mass loss.  Metallicity dependence of the IFMR is another important parameter to consider, but it will 
not be quantitatively analysed here with such incomplete metallicity information.  For our three clusters 
Pleiades, NGC 2168, and NGC 2516 with spectroscopic metallicity determinations, however, we note 
that the high-mass white dwarfs from the metal-poor NGC 2168 primarily fall above the mean
of the IFMR while the white dwarfs from the slightly metal-rich NGC 2516 fall below the mean.

Direct comparison of the NGC 2323 white dwarfs to the rest of our sample shows that they deviate 
more from the relation than 
a majority of the others.  Individually, their errors in white dwarf mass are 
large enough to make these deviations not appear significant, but both white dwarfs being 
so consistent suggests that this deviation may be telling us something.  For example, in comparison 
to the progenitors of the three comparable mass white dwarfs in other clusters (LB 1497, 
NGC 2168-LAWDS27, and NGC 2287-4), their three progenitors are all higher mass with LB 
1497 giving a $\sim$1.2 M$_\odot$ larger progenitor.  Is this indicative of mass-loss 
variations between these different clusters or remaining systematics?  For example, a possible 
explanation for this is that our assumed solar metallicity for NGC 2323 is not appropriate.   
A more metal-rich isochrone would give a younger age and help to decrease
the observed differences in progenitor masses.  Or this deviation may simply be 
the result of potential systematic differences between our adopted photometries, and in 
general this may be a cause for much of the remaining scatter in our total IFMR.  In any 
case, this new NGC 2323 white dwarf analysis further emphasizes the importance of this 
nearby, rich, and young open cluster, and of the need for spectroscopic metallicity and precise 
UBV photometry spanning from its turnoff to its faint white dwarfs.  

\vspace{1cm}
\section{Summary}

We have spectroscopically observed a sample of 10 white dwarf candidates in the young and
rich NGC 2323.  While we only found that two of these white dwarfs are consistent
with membership in NGC 2323, they are both newly discovered high-mass white dwarfs at
$\sim$1.07 M$_\odot$.  To supplement these new white dwarfs we have reanalyzed
all published, publicly available high-mass white dwarfs from the Pleiades, NGC 2516, NGC 2168,
NGC 2287, and NGC 3532 star clusters and Sirius B.  

At these higher masses, the inferred progenitor
masses are increasingly dependent on the adopted star cluster ages.  Therefore, we uniformly
analyzed available high-quality UBV photometry for these clusters.  Due to the difficulties
of analyzing younger cluster ages, we have analyzed both color-color diagrams and color-magnitude
diagrams.  The color-color diagrams provide a self-consistent
manner to not only determined a cluster reddening but to identify peculiar stars in each
cluster's turnoff.  With the cleaned turnoffs for each cluster, we can more precisely determine
a cluster age.  In general, we find cluster ages moderately older than those adopted in previous 
IFMR analyses, and we attribute this both to cluster peculiar stars affecting previous age 
determinations and the systematically older ages derived from the Y$^2$ and PARSEC isochrones.

Application of our derived white dwarf cooling ages and cluster turnoff ages gives us the
progenitor masses for each white dwarf.  The two massive white dwarfs from NGC 2323 at 1.07
M$_\odot$ both find remarkably consistent progenitor masses of 4.69 M$_\odot$.  With the addition of 
the publicly available high-mass white dwarfs, the derived IFMR is quite clean, and we find that
with Y$^2$ ages the IFMR can be well defined by a linear relation from progenitor masses of 3 to 
6.5 M$_\odot$.  In contrast, adoption of the only moderately younger cluster ages from the PARSEC 
isochrones gives a moderate turnover in IFMR slope near initial mass of 4 M$_\odot$.  In our current 
analysis we continue to adopt the Y$^2$ results, but this draws attention to the work that is needed 
on evolutionary timescales at these higher masses.  

Our IFMR also finds a significantly decreased scatter in comparison to recent IFMR work.  This 
shows that the typically observed scatter in the IFMR is likely not indicative of real variations 
in mass loss for singly evolved stars, but was the result of remaining systematics.  The IFMR 
relation is now reassuringly consistent across multiple star clusters.  Metallicity, however, 
still may play an important role, in particular at these higher masses.  To further improve the 
IFMR, not only are newly discovered high-mass white dwarfs needed, but uniform
UBV photometry and spectroscopic metallicity analyses are also needed.  This will further limit 
systematics that may still remain and give the foundation for detailed quantitative analysis of the
metallicity dependence of the IFMR.

\vspace{0.7cm}
This project was supported by the National Science Foundation (NSF) through grant AST-1211719.
This work was also supported by a NASA Keck PI Data Award, administered by the NASA Exoplanet
Science Institute. Data presented herein were obtained at the W. M. Keck Observatory from
telescope time allocated to the National Aeronautics and Space Administration through the agency’s
scientific partnership with the California Institute of Technology and the University of California.
The Observatory was made possible by the generous financial support of the W. M. Keck Foundation.
P-E.T. was supported during this project by the Alexander von Humboldt Foundation and by NASA 
through Hubble Fellowship grant HF-51329.01, awarded by the Space Telescope Science Institute, 
which is operated by the Association of Universities for Research in Astronomy, Incorporated, 
under NASA contract NAS5-26555.  This research has made use of the Keck Observatory Archive 
(KOA), which is operated by the W. M. Keck Observatory and the NASA Exoplanet Science Institute 
(NExScI), under contract with the National Aeronautics and Space Administration.  Lastly, we 
would like to thank C. Deliyannis for his helpful discussions of open cluster color-color analysis.

\end{document}